\documentclass[10pt,dvips]{article}
\usepackage{epsfig,times}

\begin{document}

\begin{center}
{\bf \Large Using cascade development universality for thin calorimeter} \\ [2mm]
{\it Gaitinov A.Sh., Ibraimova S.A., Lebedev I.A., Lebedeva A.I.} \\ [2mm]
{ Institut of Physics and Technology, Almaty, Kazakhstan}
\end{center}

\begin{abstract}
A method for measurement of energy of high-energy particles by a thin calorimeter, is presented. 
The method is based on the correlation analysis of dependence of number of secondary particles, $N_e$, at observation level and 
the relation of number of particles, $dN$, at two levels, divided by an absorber layer. 
It is shown, that use of correlation curves ($log N_e$ versus $dN$) allows to reduce essentially errors of definition of energy 
of the primary particle, which are connected with uncertainty of a primary nucleus and with fluctuations in development of cascade process.
\end{abstract}

\section{Introduction}

Calorimetric methods are the most exact for measurement of energy of high-energy nuclei and they are widely applied in experiments on accelerators and with cosmic radiation \cite{1}.

The most of methods of calorimetric measurements of energy are based on full absorption of energy of a particle in certain volume of substance.

The technical embodiment of modern ionization calorimeters can be various, but the idea remains invariable: 
the primary particle enters into dense substance (for example, iron or lead), in which numerous nuclear and electromagnetic interactions occur. 
It gives rise to a cascade of secondary particles. If depth of substance is sufficient, all kinetic energy of a primary particle will be transformed into the cascade of secondary particles, which will lose energy on ionization. 

For measurement of characteristics of the cascade the dense substance is sandwiched with special detectors. By measurement of signals from these detectors the cascade curve is formed. It represents dependence of number of particles in the cascade versus penetration depth of the cascade in calorimeter. If a maximum of a cascade curve is measured, then the energy of a primary particle can be defined. The main problem of such measurement of energy is massive installations as the calorimeter should have big enough depth for formation of the cascade curve. It considerably complicates possibilities of use of such device in the cosmic industry.

Weight reduction can be reached by using a thin calorimeter.

For definition of energy of a primary particle in a thin calorimeter the formation of the whole cascade curve is not required. In this case, a registration of its beginning is sufficient. The energy is defined by measurement of number of particles in the cascade, as number of particles on certain depth of development of the cascade is proportionally energy of a primary particle.

Thus the problem of measurement of energy of a primary particle boil down to the decision of a inverse task by simulation of development of cascade process on the basis of modern knowledge of the elementary act of interaction.

In the project the NUCLEON \cite{2} reduction of weight of equipment is reached by use of kinematic methods of definition of energy of a primary particle. This technique is based on registration of corners of scattering of the secondary particles, which are born in interaction of projectile nucleus with a target nucleus.

However using only kinematic methods leads to significant uncertainties of energy definition. Therefore the combined approach has been offered: to measure not only "width" (spatial density) of the cascade of secondary particles, but also their quantity, i.e. to unite a kinematic method with a method of a thin calorimeter. 

Authors named this technique KLEM (kinematic light - weight energy meter). 

Calculations and test experiments on the accelerator have shown, that accuracy of definition of energy will be about 50$\%$ taking into account an aprioristic spectrum of cosmic rays.

Such low accuracy is caused by essential dependence of results of energy measurement on fluctuations in development of cascade process and on mass of primary nucleus.

Influence of fluctuations in cascade development on results of measurement of energy can be reduced essentially by using correlation methods of the analysis of development of the cascade. It allows to raise accuracy of measurement of energy considerably.

The approach of correlation curves has recommended itself successfully at the analysis of extensive air showers \cite{3}. 

Modified technique, using cascade development universality for thin calorimeter, is applied in the given work.

\section{Method of correlation curves}

The method is developed and tested for the decision of a problem of measurement of energy of high-energy particles by thin calorimeter on the basis of computer calculations. 
Simulation of development of the cascade processes, initiated by primary particles of various masses and energies, has been performed on 
the basis of software package CORSIKA QGSJET \cite{4}.

In Figure 1-left the cascade curves of extensive air showers, formed by iron nuclei with various primary energies, are presented. As it is seen from the Figure, uncertainty of results of measurements of energy on the basis of a thin calorimeter first of all is connected with fluctuations in cascade development. Cascade curves essentially fluctuate and practically merge (not separable) at small values of penetration depth d. This fact does not allow to use them for definition of energy of a primary particle on the basis of a thin calorimeter, i.e. on the basis of the limited quantity of measurements on an ascending branch of a cascade curve.
\begin{figure}[t]
\begin{center}
\includegraphics*[width=0.48\textwidth,angle=0,clip]{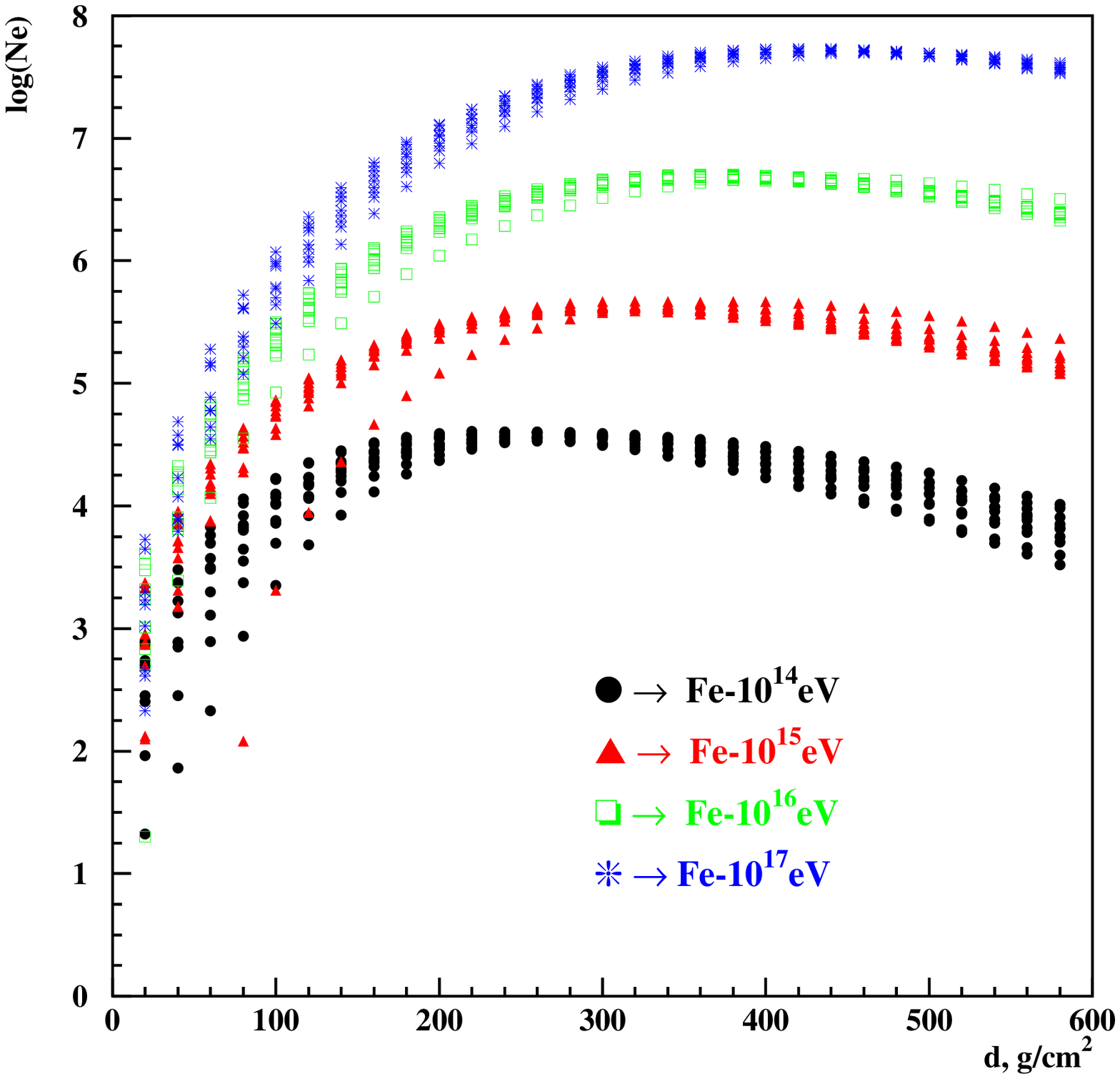}
\includegraphics*[width=0.48\textwidth,angle=0,clip]{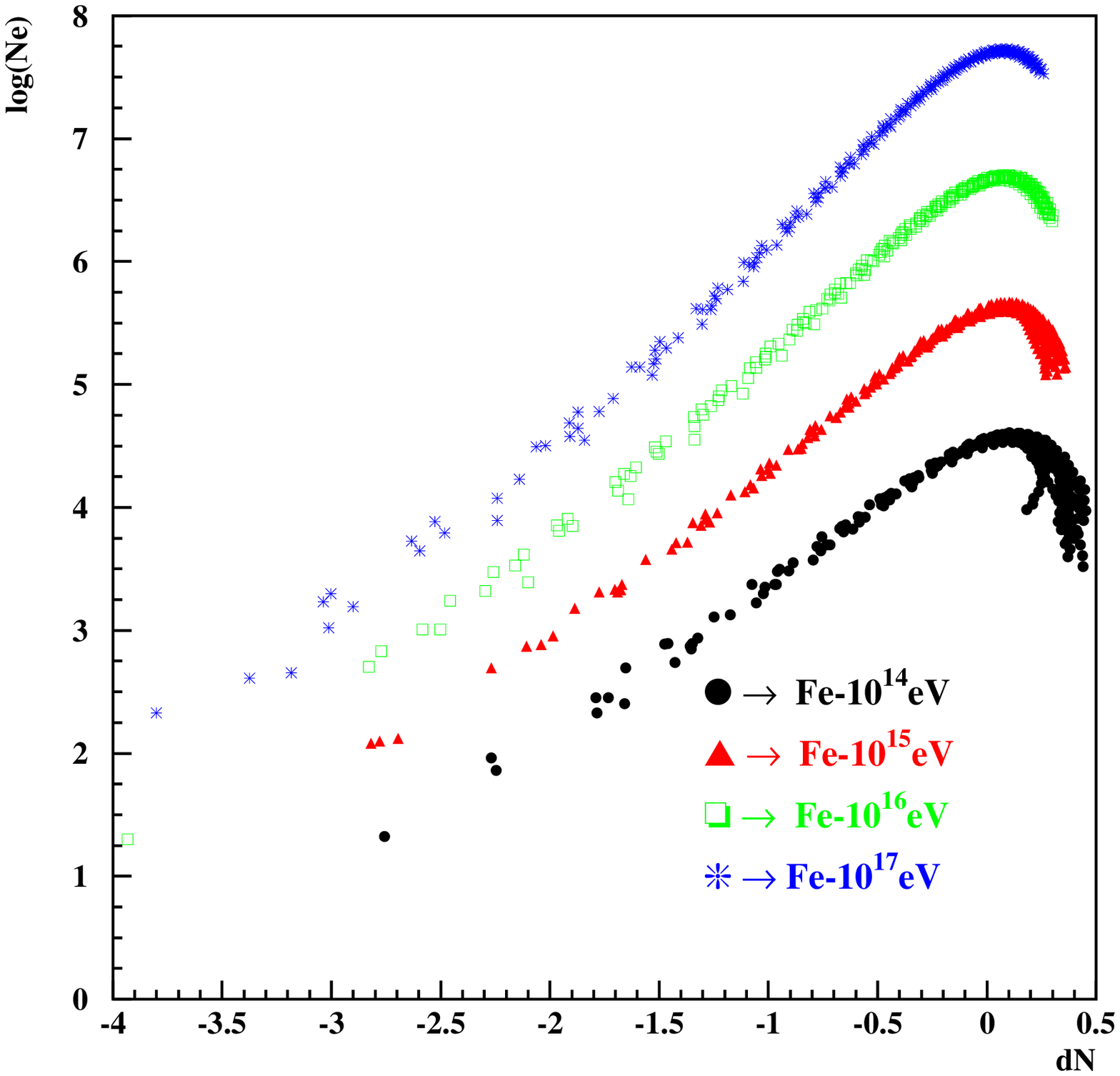}
\caption{\label{fig1} 
Cascade (left) and correlation (right) curve from interactions of iron nuclei at energies 
$10^{14}$eV, $10^{15}$eV, $10^{16}$eV, $10^{17}$eV with nuclei of atoms of air.
}
\end{center}
\end{figure}

In the technique of correlation curves some internal correlations, which are independent on fluctuations in cascade development, are analyzed. 

As a result of research of the various parameters, characterizing development of cascade process, it is revealed, that the accuracy of definition of energy can be increased essentially if to use correlation curve of dependences of particles number at observation level versus a relation of number of particles at two levels, divided by a layer of an absorber.

In Figure 1-right the given correlation curves for the same interactions, as in Figure 1-left, are presented.

Correlation curves, as it is seen from the Figure, represent more ordered picture. Fluctuations of an ascending branch of a correlation curve are not so much considerable, as in case of cascade curves. 

The second most important parameter, influencing accuracy of measurement of energy, is an uncertainty of a primary nucleus. 

In Figure 2-left average cascade curves of interaction of nuclei of iron and proton at the fixed energy $10^{16}$eV with nuclei of air atoms, are presented.

\begin{figure}[t]
\begin{center}
\includegraphics*[width=0.48\textwidth,angle=0,clip]{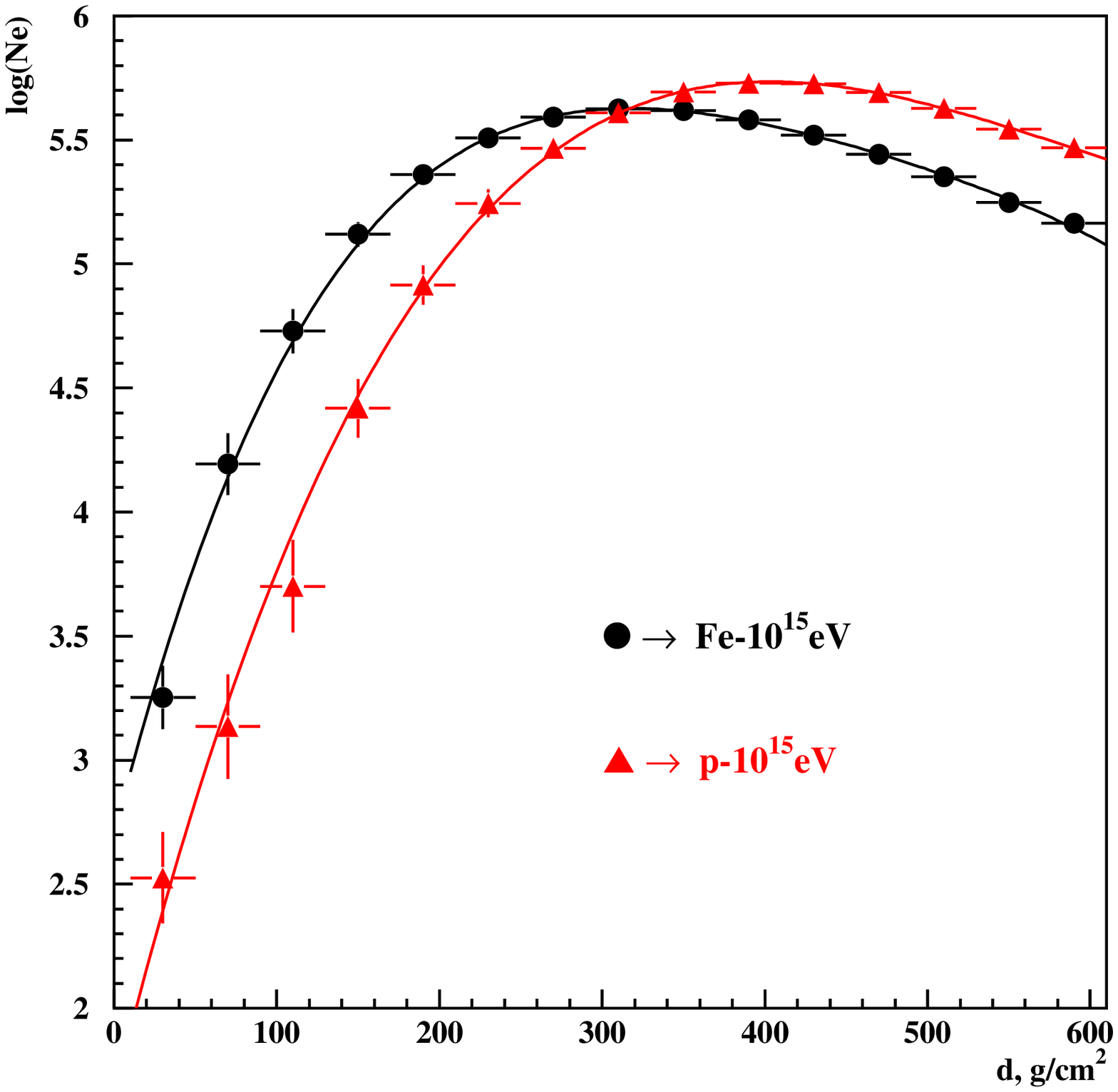}
\includegraphics*[width=0.48\textwidth,angle=0,clip]{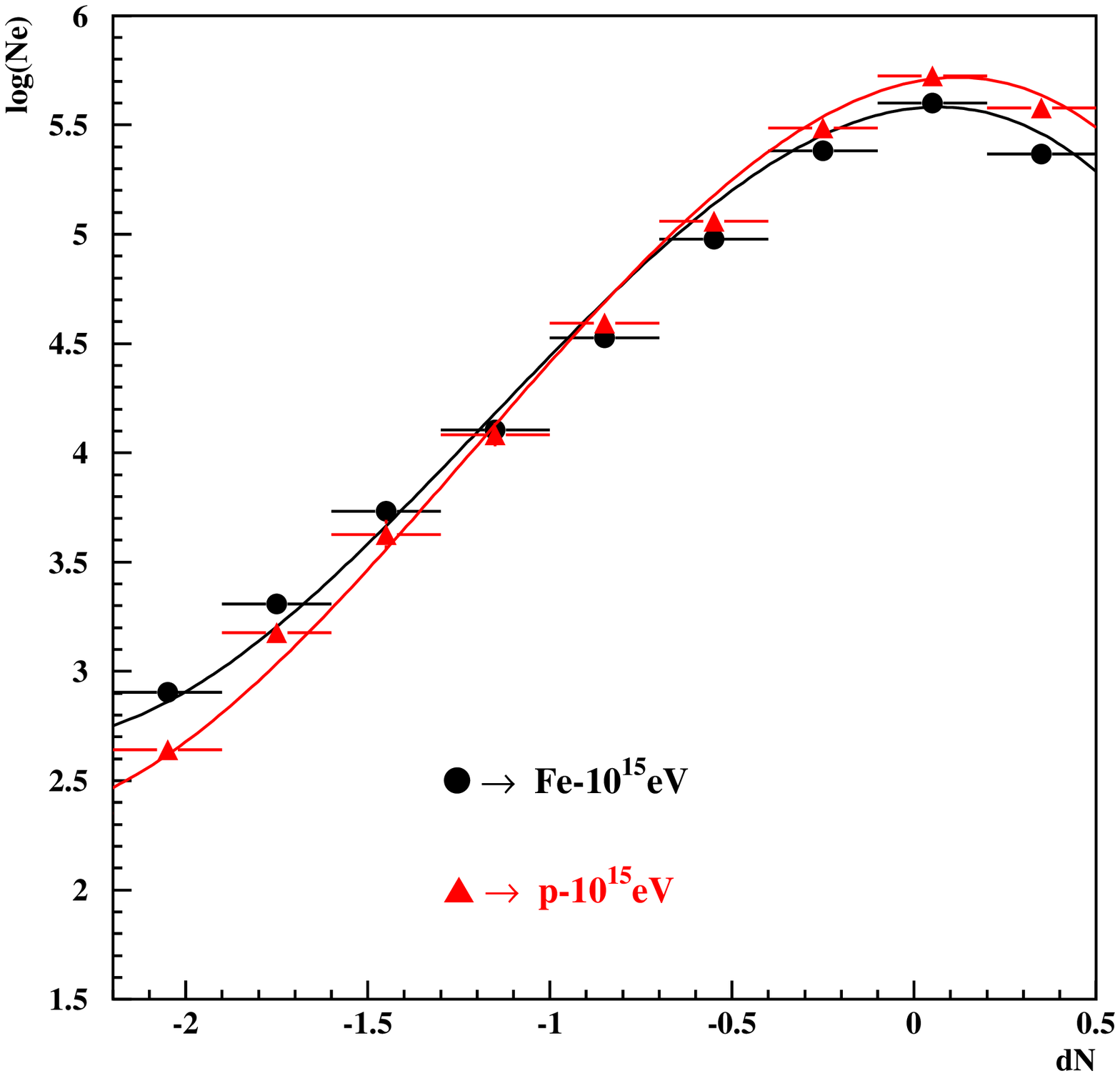}
\caption{\label{fig2} 
Average cascade (left) and correlation (right) curve of interactions of primary nuclei of iron and  proton at energy 
$10^{16}$ eV with nuclei of air atoms.
}
\end{center}
\end{figure}

The cascade curves, formed by a proton, are shifted in area of the bigger depths of penetration in comparison with Fe cascade curves. This fact leads to unequal definition of energy for various nuclei. It is connected with following.

The most of experimental groups define the primary energy $E$ on the base of measurement of number of secondary particles $N_e$ on observation level, using the dependence: 

\begin{center}
$N_e=\alpha E^\beta $
\end{center}

where $\alpha$, $\beta$ - parameters, which depend on depth of penetration and mass of a primary particle. Statistically the equation works correctly.

However, on the ascending branch of the cascade curve the energy is defined above real value for quickly developing cascades, and below real energy for slowly developing cascades. 

This fact leads to the underestimated value of energy of proton cascades and to the overestimated value for Fe cascades.

Using correlation curves allows to reduce essentially energy definition errors, connected with uncertainty of a primary nucleus. 
Average correlation curves are presented in Figure 2-right. As it is seen from the Figure, correlation curves practically coincide for different nuclei. Maximum points of proton and iron correlation curves are shifted in one point $dN=0$, which is independent from depth of penetration. 

Summing up the given section once again we will underline the following: accuracy of definition of energy on the basis of a thin calorimeter can be increased essentially if to use correlation curve dependences of number of particles at observation level versus a difference of number of particles at two levels, divided by an absorber layer.

The following important question concerns a choice of an optimum thickness of a layer of an absorber. 

\section{Discretization parametres of cascade curves}
 
Discretization parametres of cascade curves are important value for definition of optimum density of substance and minimisation of number of layers of a thin calorimeter. In Figure 3 correlation curves with various values of a thickness of an absorber $dN$, are presented.

\begin{figure}[t]
\begin{center}
\includegraphics*[width=0.48\textwidth,angle=0,clip]{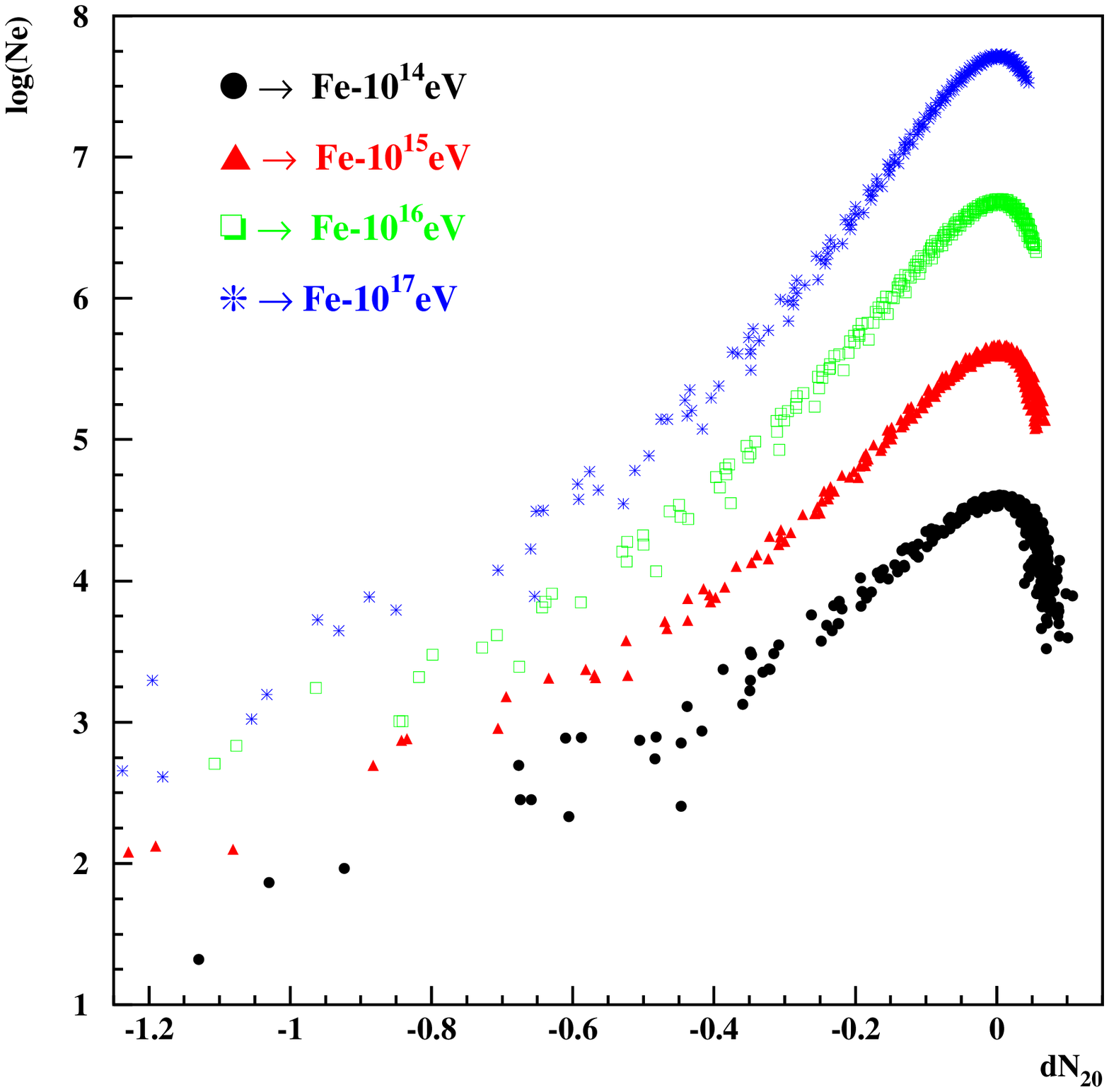}
\includegraphics*[width=0.48\textwidth,angle=0,clip]{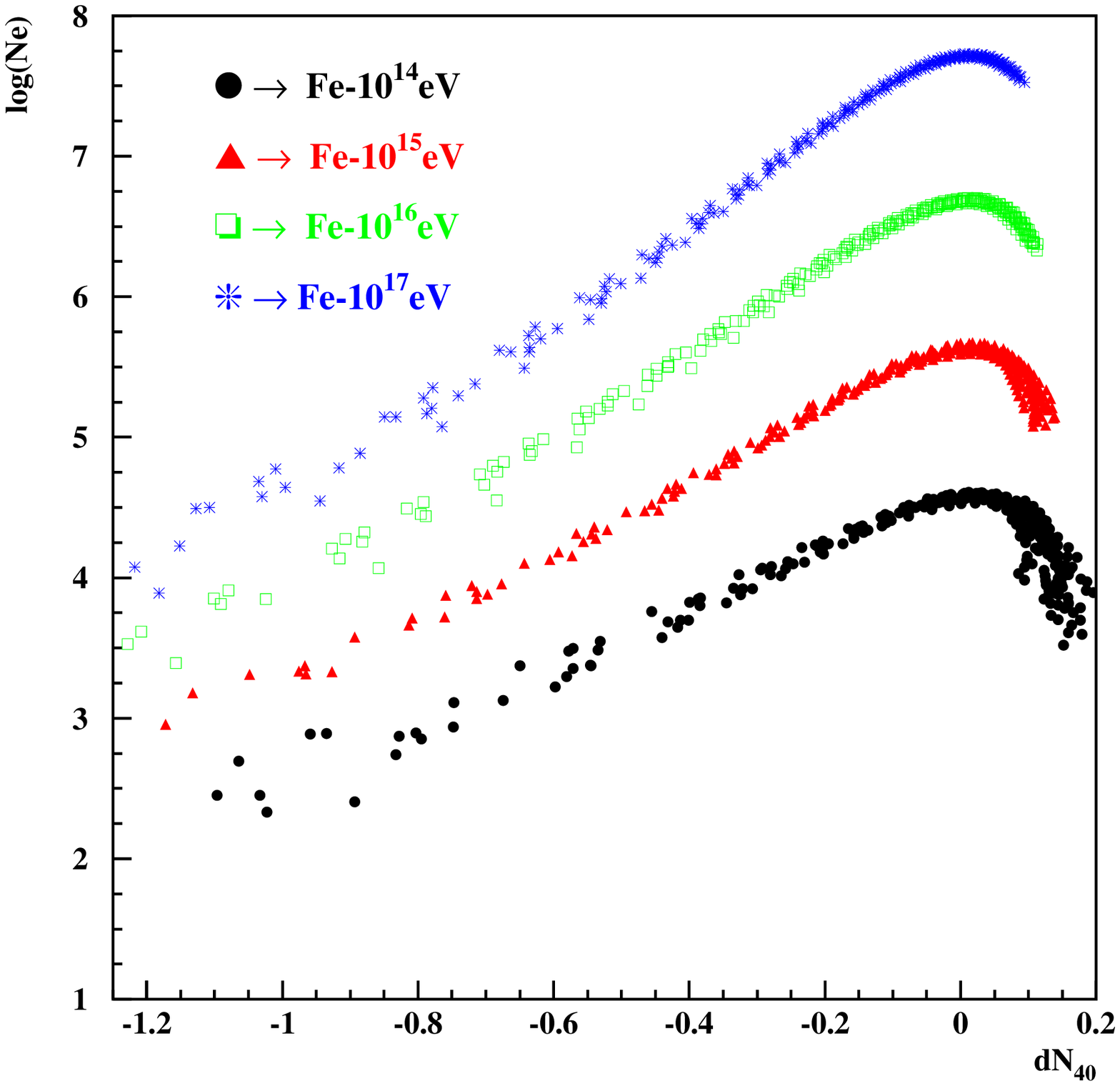}
\includegraphics*[width=0.48\textwidth,angle=0,clip]{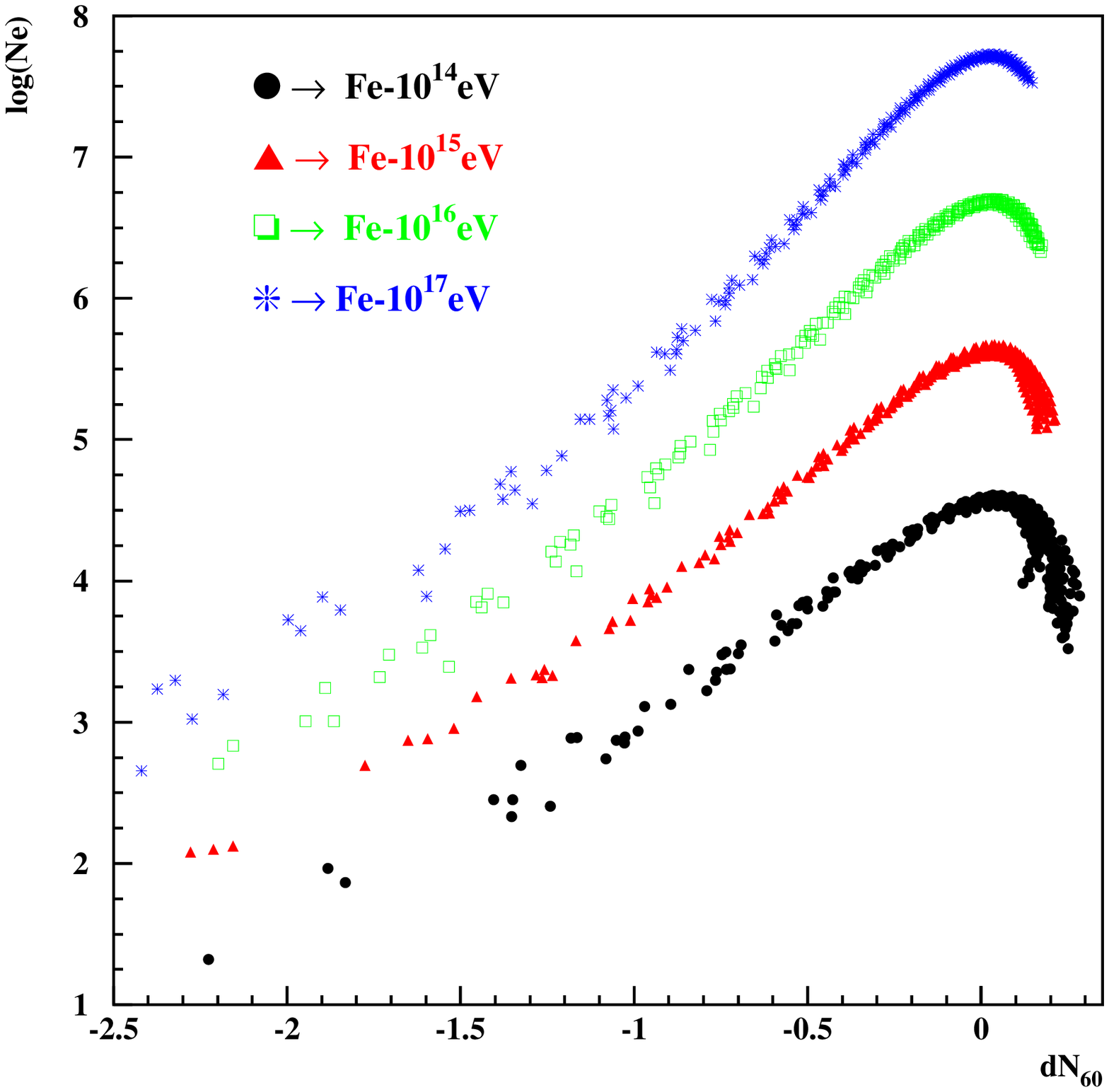}
\includegraphics*[width=0.48\textwidth,angle=0,clip]{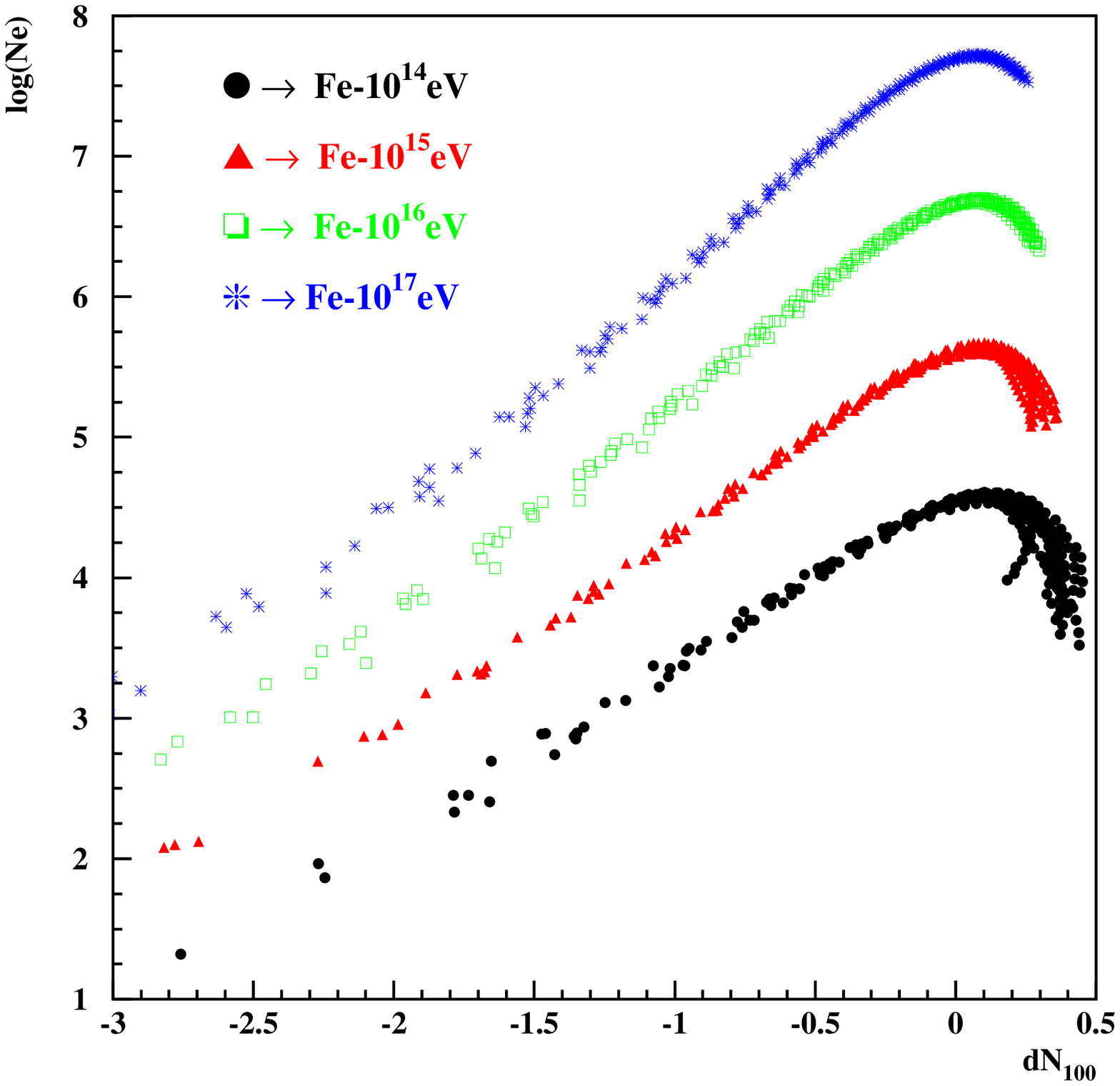}
\caption{\label{fig3} 
Correlation curves (Fe $10^{14}$eV, $10^{15}$eV, $10^{16}$eV, $10^{17}$eV) with different values of a thickness of a layer 
($dN$= 20, 40, 60, 100 $g/sm^2$)}
\end{center}
\end{figure}

As it is seen from the Figure, the increase in layer thickness of an absorber leads to increase in accuracy of definition of energy. However the increase of the thickness increases weight of installation.

Thus, the choice of a thickness of a layer depends on conditions of concrete experiment.

\section{Conclusion}

The technique of measurement of energy of primary cosmic particles on the basis of correlation research of development of cascade process in consistently located layers of a thin calorimeter, is developed.

The given technique is based on the correlation analysis of dependence of number of secondary particles at observation level and the relation of number of particles at two levels divided by a layer of an absorber. It is shown, that use of correlation curves allows to reduce essentially errors of definition of energy of the primary particle, connected with uncertainty of a primary nucleus and fluctuations of development of cascade process. 

Further for possible technical realisation of the project of a thin calorimeter it will be necessary to solve problems connected with calculation of installation response, optimisation of the information gathering, etc.  

However, now the basic result is received: on the basis of  computer simulation the correlation parameters, which allow to define the energy of a primary nuclei on an ascending branch of a cascade curve, are found out. 

Work is supported by grant of MES RK N1276/GF2.

\end{document}